\documentstyle [12pt, epsf] {article}

\catcode`@=12
\newcommand{\be}{\begin{equation}}
\newcommand{\ee}{\end{equation}}
\newcommand{\bea}{\begin{eqnarray}}
\newcommand{\eea}{\end{eqnarray}}
\newcommand{\nn}{\nonumber}
\def\ds{\displaystyle}
\def\s1{\hat s}
\def\ve{\varepsilon}
\def\rar{\rightarrow}
\def\ar{&+& \!\!\!}
\def\ek{&-& \!\!\!}
\def\es{&=& \!\!\!}
\begin{document}
\vspace{0.5in}
\oddsidemargin -.1 in
\newcount\sectionnumber
\sectionnumber=0
\def\lsim{\mathrel{\vcenter{\hbox{$<$}\nointerlineskip\hbox{$\sim$}}}}
\thispagestyle{empty}

\vskip.5truecm
\vspace*{0.5cm}

\begin{center}
{\Large \bf \centerline{Study of FCNC mediated  rare $B_s$ decays in}} 
{\Large \bf \centerline{  a single  universal extra dimension scenario}}
 \vspace*{0.5cm} {R. Mohanta$^1$ and A.K Giri$^2$}
\vskip0.3cm {\it $^1$ School of Physics, University of Hyderabad,
 Hyderabad
500 046, India\\
$^2$ Department of Physics, Punjabi University, Patiala 147 002, India}\\
\vskip0.5cm
\bigskip
\begin{abstract}
We study the rare semileptonic and radiative leptonic
$B_s$ decays  in the universal extra 
dimension model. In this scenario, with a single extra
dimension,  there exists  only one new parameter beyond those
of the standard model, which is the inverse of the
compactification radius $R$. We find that  with the additional
contributions due to the KK modes the branching ratios of the rare
 $B_s$ decays are enhanced from their 
corresponding standard model values and the zero point of the
forward backward asymmetries are shifted towards the left.
\end{abstract}
\end{center}

\thispagestyle{empty}
\newpage

\section{Introduction}

Although the standard model (SM) of electroweak 
interaction is very successful
in explaining the observed data so far, but still it is believed that
there must exist some new physics beyond the SM, whose true nature is 
not yet well-known. Therefore, intensive search for physics beyond the 
SM is now being performed in various areas of particle physics. In this 
context the $B$ system can also be used as a complementary 
probe for new physics.
Unfortunately, we have not been able to see any clear indication of
physics beyond the SM in the currently running $B$-factories (SLAC
and KEK). Nevertheless, there appears to be some kind of deviation
in some $b\to s$ penguin induced transitions like the deviation
in the measurement of sin2$\beta$ in $B_d \to \phi K_S $ and also in
some related processes, polarization anomaly in $B\to \phi K^*$
 and deviation of branching ratios from the SM expectation 
in some rare $B$ decays, etc. \cite{ref1}.
But, the present situation is that it is too early to substantiate 
or rule out the existence of new physics in the b-sector.

One of the important ways to look for new physics in the b-sector is
the analysis of rare $B$ decay modes, which are induced by the
flavor changing neutral current (FCNC) transitions. The 
FCNC transitions generally arise at the loop level in the SM, 
and thus provide an excellent
testing ground for new physics. Therefore, it is very important to
study the FCNC processes, both theoretically and experimentally,
as these decays can provide a sensitive test for the investigation
of the gauge structure of the SM at the loop level. Concerning the
semileptonic $B$ decays, $B\to X_s~ l^+l^-$ ($X_s= K, K^*,~ l=e, \mu,
\tau$) are a class of decays having both theoretical and
experimental importance. At the quark level, these decays proceed
through the FCNC transition $ b\to s l^+ l^-$, which occur only through
loops in the SM, and therefore, these decays constitute a quite 
suitable tool of looking for new physics. 
Moreover, the dileptons present in these processes allow us to
formulate many observables which can serve as a testing ground to
decipher the presence of new physics. In this context,
the study of the rare semileptonic $B_s$ decay modes, 
$B_s \to \phi l^+ l^-$, which are 
induced by the same quark level transition, i.e.,  $b\to s l^+l^-$, 
might be worth exploring. These decay modes may provide us additional 
information towards the quest for
the existence of new physics beyond the SM and therefore deserve serious
attention, both theoretically and experimentally. Since at the
quark level they are induced by the same mechanism, so we can also
independently test our understanding of the quark-hadron dynamics
and also study some CP violation parameters with the help of
rare $B_s$ decays, apart from corroborating the finding of the
$B$-meson sector. These decay modes are studied in the standard model
\cite{ref2} and in the two Higgs doublet model \cite{ref3}. Recently,
the D0 collaboration \cite{ref4} has  reported a more
stringent  upper limit  
on the branching ratio of $B_s^0 \to \phi \mu^+ \mu^-$
mode as Br$(B_s^0 \to \phi \mu^+ \mu^-) 
< 4.1 \times 10^{-6}$.

Another sensitive  process to look for new physics
is the radiative dileptonic decay mode $B_s^0 \to l^+ l^- \gamma$. 
In contrast to the pure leptonic decays $B_s^0 \to l^+ l^-$,
these modes are free
from helicity suppression due to the emission of a photon in addition
to the lepton pair. Thus, the branching ratios of these processes 
are larger than those  for the pure leptonic modes despite of
an additional $\alpha$ suppression. These modes are also studied 
in the SM \cite{eilam} and some beyond the SM scenarios \cite{aliev}.

New physics effects manifest themselves in these rare $B_s$
decays in two different ways, either through new contribution to
the Wilson coefficients or through the new structure in the
effective Hamiltonian, which are absent in the SM.
There are many variants of possible extensions to the SM exist 
in the literature but the models with extra dimensions have received
considerable attention in recent years. An elegant beyond the SM scenario 
with extra dimensions is being considered to solve the classical problems
of particle physics, mainly the hierarchy problem, gauge coupling
unification, neutrino mass generation, fermion mass hierarchies etc.
In fact, there are several models that exist in the literature 
which take into account the effect of
extra dimensions and they differ from one
another depending on the number of extra dimensions, the geometry of 
space-time, the compactification manifold, which particles can go into the
extra-dimensions and which cannot etc. In the case of the scenario with 
Universal Extra Dimensions (UED), proposed by 
Appelquist-Cheng-Dobrescu (ACD) \cite{acd}, 
all the fields are allowed to propagate in all
available dimensions. 
In its simplest version the single
extra dimension is taken to be the fifth dimension, that is, $x_5=y$. 
This is in fact compactified on the orbifold $S^1/Z_2$, i.e., on a circle
of radius  $R$ and runs from 0 to $2 \pi R$,  with $y=0$ and  $y=\pi R$
are the fixed points of the orbifold. Hence a field $F(x,y)$, where $x$ denotes
the usual $3+1$ dimension, would be a periodic function of $y$ and can
be represented as
\bea
F(x,y)=\sum_{n=-\infty}^{\infty} F_n(x) e^{i n \cdot y}\;.
\eea
Under the parity transformation $P_5: y \to -y$, fields which
exist in the SM are even and their zero modes in the KK expansion 
are interpreted as ordinary SM fields. On the other hand, fields
absent in the SM are odd under $P_5$, so they do
not have zero modes.
Compactification of the extra dimension  leads to the appearance of
Kaluza-Klein (KK) partners of the SM fields as well as KK modes 
without corresponding SM partners.
Thus, in the particle spectrum of ACD model, in addition to the 
ordinary particles of the SM denoted 
as zero mode ($n=0$), there are infinite towers of Kaluza-Klein 
modes ($n \geq 1$). There is one such tower of each SM boson, two
for each SM fermion, and for the physical scalars $a_{(n)}^{0,\pm}$,
no zero mode exists. The masses of the bosonic $KK$ modes are given as
$m_{(n)}^2=m_0^2+n^2/R^2$, with $m_0$ being the mass of the zero mode.
The important features of the ACD model are: (i) there is only one
additional free parameter with respect to the SM, i.e., $1/R$,
the  inverse of the compactification radius (ii) the KK parity is conserved,
which implies that the KK modes do not contribute at tree level 
for the low energy interaction processes and (iii) the lightest KK particle
must be stable.

The implications of physics with UED are being examined with the data
from accelerator experiments,
for example, from Tevatron experiments the bound on  the inverse
compactification radius is found to be about $1/R\ge 300$ GeV.
Analysis of the anomalous magnetic moment  and $Z \rar \bar{b}
b$ vertex \cite{agashe} also lead to the bound $1/R\ge 300$ GeV.
Possible manifestation of ACD model in the $K_L- K_S$ mass difference,
$B^0-\bar{B}^0$ mixing, rare decays of $K$ and $B$ mesons
 and the kaon CP violation parameter $\varepsilon^\prime/\varepsilon$ are
comprehensively investigated in \cite{buras1}. Exclusive 
$B \rar K^\ast \ell^+ \ell^-$, $B \rar K^\ast \bar{\nu} \nu$ and $B \rar
K^\ast \gamma$ decays {\cite{fazio} and rare semileptonic
$\Lambda_b$ decays \cite{lam} are also studied in the framework 
of the UED scenario.

In this work, we would like to study the implications of the scenario
with universal
extra dimensions in the semileptonic
rare decay modes  $B_s^0 \to \phi~l^+ l^- $, and the
radiative dileptonic decays $B_s^0 \to l^+ l^- \gamma $.
As discussed above, since these decays are very sensitive to new physics,
these studies may provide an indirect way to constrain the new physics 
parameter $R$. The paper is organized as follows. In section II, we 
present the effective Hamiltonian for these decays in the ACD model.
The decay rates and forward backward rate asymmetries for
the mode $B_s^0 \to \phi l^+ l^- $ and $B_s^0 \to l^+ l^- \gamma$
are considered in section III.
Section IV contains our
conclusion.  

\section {Effective Hamiltonian}
 The effective Hamiltonian describing the 
$b \to s l^+ l^- $ transition 
in the SM can be written as \cite{buras2}
\bea
{\cal H}_{eff} &= &\frac{ G_F ~\alpha}{\sqrt 2 \pi} V_{tb} V_{ts}^*\Big[ 
C_9^{eff}(\bar s \gamma_\mu L b)(\bar l \gamma^\mu l) \nn\\
&+& C_{10}(\bar s \gamma_\mu L b)(\bar l \gamma^\mu \gamma_5 l) -2 C_7^{eff}
 m_b(\bar s i \sigma_{\mu \nu} \frac{q^\nu}{q^2} R b)
(\bar l \gamma^\mu l) \Big]\;,\label{ham}
\eea
where $q$ is the momentum transferred to the lepton pair, given as
$q=p_-+p_+$, with $p_-$ and $p_+$ are the momenta of the leptons $l^-$ 
and $l^+$ respectively. $L,R=(1 \pm \gamma_5)/2$ and $C_i$'s are the Wilson
coefficients evaluated at the $b$ quark mass scale.
The coefficient $C_9^{eff}$ has a perturbative part  and a 
resonance part which comes
from the long distance effects due to the conversion of the real
$c \bar c$ into the lepton pair $l^+ l^-$. Hence, $C_9^{eff}$ can be 
written as 
\be
C_9^{eff}=C_9+Y(s)+C_9^{res}\;,
\ee
where $s=q^2$ and the function $Y(s)$ denotes the perturbative part coming 
from one loop matrix elements  of the four quark operators and
is given in Ref. \cite{buras3}.
The long distance resonance effect is given as \cite{res}
\bea
C_9^{res}= \frac{3 \pi}{\alpha^2}(3 C_1+C_2+3C_3+C_4+3C_5+C_6)\sum_{V_i=J/\psi,
\psi^\prime} \kappa\frac{m_{V_i} \Gamma(V_i \to l^+ l^-)}{m_{V_i}^2 -s
-i m_{V_i}\Gamma_{V_i}}\;,
\eea
where the phenomenological parameter $\kappa$ is taken to be 2.3, so as to
reproduce the correct branching ratio  $ {\cal B}(B \to J/\psi K^*
\to K^* l^+ l^-)={\cal B}(B \to J/\psi K^*){\cal B}(J/\psi \to l^+ l^-)$.

Now we will analyze the KK contributions to the above decay modes. Since
there is no tree level
contribution of KK modes in low energy processes, the new contributions, 
therefore, only come from the loop diagrams with internal KK modes.  
It should be noted here that there does not
appear any new operator in the ACD model, and therefore, new effects are
implemented by modifying the Wilson coefficients existing in the SM, if we
neglect the contributions of the scalar fields, which are indeed very small. 
Thus, the modified Wilson coefficients are given as \cite{buras1}
\bea
C_1(M_W)&=&\frac{11}{2} \frac{\alpha_s(M_W)}{4 \pi}\;,~~~~~C_2(M_W)=
1-\frac{11}{6} \frac{\alpha_s(M_W)}{4 \pi}\;\nn\\
C_3(M_W)&=&-3C_4(M_W)=C_5(M_W)=-3C_6(M_W)=- \frac{\alpha_s}{24 \pi} \tilde
E(x_t, 1/R)\;,\nn\\
C_7(M_W)&=&- \frac{1}{2} \tilde
D'(x_t, 1/R)\;,~~~C_8(M_W)=- \frac{1}{2} 
E'(x_t, 1/R)\;.
\eea
The functions $\tilde E(x_t, 1/R)$, $E'(x_t, 1/R)$ and $
D(x_t, 1/R)$  are given by
\be
F(x_t, 1/R)=F_0(x_t)+\sum_{n=1}^\infty F_n(x_t, x_n)\;,~~~F=\tilde 
E, E',D'\;,
\ee 
where $x_t=m_t^2/M_W^2$, $x_n=m_n^2/M_W^2$ and $m_n=n/R$.
The loop functions $F_0(x_t)$ and  $F_n(x_t, x_n)$ are given as \cite{buras1}
\bea
E_0(x_t) &=& -\frac{2}{3} \ln x_t + \frac{x_t^2(15-16 x_t+4 x_t^2)}{
6(1-x_t)^4}\ln x_t+ \frac{x_t(18-11 x_t-x_t^2)}{12(1-x_t)^3}\;, \nn\\ \nn\\
D^\prime_0(x_t) &=& - \frac{(8 x_t^3+5 x_t^2-7 x_t)}{12 (1-x_t)^3}
+ \frac{x_t^2(2-3 x_t)}{2(1-x_t)^4}\ln x_t~, \nn \\ \nn \\
E^\prime_0(x_t) &=& - \frac{x_t(x_t^2-5 x_t-2)}{4 (1-x_t)^3} +
\frac{3}{2}\frac{ x_t^2}{ (1-x_t)^4}\ln x_t~, \nn \\ \nn \\
E_n(x_t,x_n) &=&  -\frac{x_t[35+ 8 x_t-19 x_t^2+6 
x_n^2(10-9 x_t+3 x_t^2) +3 x_n (53-58 x_t+21 x_t^2)]}
{36 (x_t-1)^3 }\nn \\
\ek \frac{1}{2} (1+x_n)(-2+3 x_n+ x_n^2) \ln \frac{x_n}{1+ x_n} \nn \\
\ar \frac{(1+x_n) (-6+19 x_t-9 x_t^2 +x_n^2(3+x_t)
+x_n(9-4 x_t+3 x_t^2))}{6 (x_t-1)^4} \ln \frac{x_n+x_t}{1+x_n}~, \nn\\ \nn \\
D^\prime_n(x_t,x_n) &=&  \frac{x_t[-37+44 x_t+17 x_t^2+6 
x_n^2(10-9 x_t+3 x_t^2) -3 x_n (21-54 x_t+17 x_t^2)]}
{36 (x_t-1)^3 }\nn \\
\ar \frac{x_n(2-7 x_n+3 x_n^2)}{6} \ln \frac{x_n}{1+ x_n} \nn \\
\ek \frac{ (-2+x_n+3 x_t)[x_t+3 x_t^2+x_n^2(3+x_t) - 
x_n(1+  (-10+x_t) x_t)] }
{6 (x_t-1)^4} \ln \frac{x_n+x_t}{1+x_n}~, \nn\\ \nn \\
E^\prime_n(x_t,x_n)\es \frac{x_t[-17-8 x_t+x_t^2-3 x_n(21-6 x_t+x_t^2)-6
x_n^2(10-9 x_t+3 x_t^2)]}{12 (x_t-1)^3} \nn \\
\ek \frac{1}{2} x_n(1+x_n)(-1+3 x_n)\ln \frac{x_n}{1+ x_n} \nn \\
\ar \frac{(1+x_n) [x_t+3 x_t^2+x_n^2(3+x_t)-x_n(1  +(-10+ x_t) x_t )] }
{2(x_t-1)^4} \ln \frac{x_n+x_t}{1+x_n}~.
\eea
The functions with and without $x_n$ correspond to the KK excitation
and SM contributions, respectively. The summations
are carried out using the prescription presented in
\cite{buras1}.

After obtaining the Wilson coefficients at the scale $M_W$, we have to
run these coefficients $C_i(M_W)$ down to the scale $\mu\sim m_b$.
The Wilson coefficients 
$C_{i=1, \cdots 6}(\mu \sim m_b)$ at low energy  can be obtained from 
the corresponding $C_{i=1, \cdots 6}(M_W)$ ones
by using the Renormalization Group (RG) equation, as discussed in
Ref. \cite{buras2}, as
\be
{\bf C}(\mu) ={\bf U}_5(\mu, M_W) {\bf C}(M_W)\;,
\ee
where ${\bf C}$ is the $6 \times 1$ column vector of the
Wilson coefficients and
${\bf U}_5(\mu, M_W)$ is the five-flavor $6 \times 6$ evolution matrix.
In the next-to-leading order (NLO), ${\bf U}_5(\mu, M_W)$ is given by
\be
{\bf U}_5(\mu, M_W)=\left (1+\frac{\alpha_s(\mu)}{4 \pi} {\bf J} \right )
{\bf U}_5^{(0)}(\mu, M_W)\left (1-\frac{\alpha_s(M_W)}{4 \pi} {\bf J} 
\right )\;,
\ee
where ${\bf U}_5^{(0)}(\mu, M_W)$ is the leading order (LO)
evolution matrix and ${\bf J}$ denotes the NLO corrections to the evolution. 
The explicit forms of ${\bf U}_5(\mu, M_W)$ and 
${\bf J}$ are given in Ref. \cite{buras2}. 
For the coefficient $C_7$ we use the RG running as
\bea
C_7^{eff}(\mu_b) &=& \eta^{\frac{16}{23}}
C_7^{}(\mu_W)+ \frac{8}{3} \left( \eta^{\frac{14}{23}} 
-\eta^{\frac{16}{23}} \right) C_8^{}(\mu_W)+C_2^{}(\mu_W) \sum_{i=1}^8
h_i \eta^{a_i}~, 
\eea
where 
$\eta = \frac{\alpha_s(\mu_W)} {\alpha_s(\mu_b)}$, and 
the coefficients $a_i$ and $h_i$ are as given in Ref. \cite{buras1}.

The Wilson coefficient $C_9$
in the ACD model and in the NDR scheme is 
\bea
C_9(\mu_b)=P_0^{NDR}+{Y(x_t,1/R) \over \sin^2 \theta_W} -4
Z(x_t,1/R)+P_E E(x_t,1/R)~, 
\eea 
where $P_0^{NDR}=2.60 \pm 0.25$
\cite{buras1} and the last term is numerically negligible.
The functions $Y(x_t,1/R)$ and $Z(x_t,1/R)$ are defined as:
\bea 
Y(x_t,1/R) &=& Y_0(x_t)+\sum_{n=1}^\infty C_n(x_t,x_n)~, \nn \\
Z(x_t,1/R) &=& Z_0(x_t)+\sum_{n=1}^\infty C_n(x_t,x_n)~, 
\eea 
with
\bea 
Y_0(x_t) &=& \frac{x_t}{8} \left[ \frac{x_t -4}{x_t -1}+
\frac{3 x_t}{(x_t-1)^2} \ln x_t \right]~,\nn \\ \nn \\
Z_0(x_t) \es \frac{18 x_t^4-163 x_t^3+259 x_t^2 -108
x_t}{144 (x_t-1)^3} \nn \\
\ar \left[\frac{32 x_t^4-38 x_t^3-15 x_t^2+18 x_t}{72
(x_t-1)^4} - \frac{1}{9}\right] \ln x_t\;, \\ \nn\\
C_n(x_t,x_n) &=& \frac{x_t}{8 (x_t-1)^2} \left[x_t^2-8 x_t+7+(3 +3
x_t+7 x_n-x_t x_n)\ln \frac{x_t+x_n}{1+x_n}\right]~. 
\eea 
Finally, the Wilson coefficient $C_{10}$, which is scale independent, 
is given by
\bea 
C_{10}= - \frac{Y(x_t,1/R)}{\sin^2 \theta_W}~.
\eea

\section{Decay rates and forward backward asymmetries}
\subsection{$B_s \to \phi~ l^+ l^- $ process}

The semileptonic rare $B$ decays are very well studied in the 
literature, both in the SM and in various extensions of it.
Therefore, we shall simply present here the expressions for the 
dilepton mass spectrum and the forward backward asymmetry
parameters.
For the $B_{s} \to \phi l^+ l^-$ process, the hadronic
matrix elements are given as
\bea
&&\langle \phi(k, \varepsilon)|(V-A)_\mu |B_s(P) \rangle =
 \epsilon_{\mu \nu \alpha \beta} \varepsilon^{* \nu} P^\alpha k^\beta
\frac{2 V(q^2)}{m_{B}+m_\phi}-i \varepsilon_\mu^*(m_{B}+m_\phi)
A_1(q^2)\nn\\
&&~~~~~~~~~~~~+
i(P+k)_\mu (\varepsilon^* q)\frac{A_2(q^2)}{m_{B}+m_\phi}
+i q_\mu(\varepsilon^*  q) \frac{2 m_\phi}{q^2}\Big[A_3(q^2)-A_0(q^2)
\Big]\;,
\nn\\ \nn\\
&&\langle \phi (k, \varepsilon)|\bar s \sigma_{\mu \nu} {q^\nu}(1 +
\gamma_5) b |B(P) \rangle  =
i \epsilon_{\mu \nu \alpha \beta} \varepsilon^{* \nu} P^\alpha k^\beta
2 T_1(q^2)+ \Big[\varepsilon_\mu^*(m_{B}^2-m_\phi^2)\nn\\ 
&&~~~~~~~~
 -
(\varepsilon^*  q)(P+k)_\mu\Big]T_2(q^2)
+  (\varepsilon^*  q)\Big[q_\mu -
\frac{q^2}{m_{B}^2-m_\phi^2}(P+k)_\mu \Big]T_3(q^2)\;,\label{vf}
\eea
where $V$ and $A$ denote the vector and axial vector currents,
$A_0,A_1,A_2,A_3,V,T_1,T_2$ and $T_3$ are the relevant form factors
and $m_B$ denotes the mass of $B_s$ meson.
Thus, with eqs. (\ref{ham}) and (\ref{vf})
the transition amplitude for $B_s \to \phi l^+ l^-$ is given as 
\bea
{\cal M}(B_s  \to  \phi~ l^+ l^-) & = & \frac{G_F \alpha}
{2 \sqrt 2 \pi} V_{tb} V_{ts}^* \biggr\{ \bar l \gamma^\mu l \Big[
2 A \epsilon_{\mu \nu \alpha \beta}\varepsilon^{* \nu} k^\alpha P^\beta 
+iB \varepsilon_\mu^*\nn\\
&- & i C (P+k)_\mu (\varepsilon^* q)
-  iD (\varepsilon^*  q)
q_\mu \Big]
+ \bar l \gamma^\mu \gamma_5 l \Big[ 2E \epsilon_{\mu \nu \alpha \beta}
\varepsilon^{* \nu} k^\alpha P^\beta \nn\\
&+ &iF \varepsilon_\mu^*
-  i G  (\varepsilon^*  q)(P+k)_\mu -iH (\varepsilon^*  q)
q_\mu \Big]\biggr\},\label{phi}
\eea
where the parameters $A,B, \cdots H$ are given as \cite{ref3} 
\bea
A &=& C_9^{eff} \frac{V(q^2)}{m_B + m_\phi}+ 
4 \frac{m_b}{q^2} C_7^{eff} T_1(q^2)\;, \nn\\
B &=& (m_B + m_\phi) \left (
C_9^{eff} A_1(q^2) + 4 \frac{m_b}{q^2}(m_B -m_\phi) C_7^{eff} T_2(q^2) 
\right )\;,\nn\\
C &=& C_9^{eff} \frac{A_2(q^2)}{m_B+m_\phi} +4 \frac{m_b}{q^2} C_7^{eff} 
\left (T_2(q^2) + \frac{q^2}{m_B^2-m_\phi^2} T_3(q^2) \right )\;,\nn\\
D &=& 2C_9^{eff} \frac{m_\phi}{q^2}\Big(A_3(q^2)-A_0(q^2)
\Big)-4 C_7^{eff} \frac{m_b}
{q^2}T_3(q^2) \;,\nn\\
E&=& C_{10} \frac{V(q^2)}{m_B+m_\phi}\;,~~~~~
F = C_{10}(m_B +m_\phi) A_1(q^2)\;,\nn\\
G&=& C_{10} \frac{A_2(q^2)}{m_B+m_\phi}\;,~~~~
H= 2 C_{10} \frac{m_\phi}{q^2}\Big(A_3(q^2)-A_0(q^2)\Big)\;.
\eea
The differential decay rate is given as
\be
\frac{d \Gamma}{d s}=\frac{G_F^2 \alpha^2}{2^{12} \pi^5 m_B} 
|V_{tb}V_{ts}^*|^2
\lambda^{1/2} (1, r_\phi, \hat s)~ v_l ~ \Delta\;,\label{lp1}
\ee
where $\hat s=q^2/m_B^2$, $r_\phi=m_\phi^2/m_B^2$, $v_l=
\sqrt{1-4m_l^2/s}$,
$\lambda \equiv \lambda(1,r_\phi,\hat s)$, is the triangle function and
\bea
\Delta &=& \frac{8}{3} \lambda m_B^6 \hat s\Big((3-v_l^2)|A|^2+2 v_l^2 
|E|^2 \Big)
+ \frac{1}{r_\phi}\lambda m_B^4 \biggr[ \frac{1}{3} \lambda m_B^2
(3-v_l^2)|C|^2\nn\\
&+& m_B^2 \hat{s}^2 (1-v_l^2)|H|^2+ \frac{2}{3} \Big[ (3-v_l^2)
(r_\phi+\hat s-1)
-3 \hat s (1-v_l^2)\Big]Re[F G^*]\nn\\
&-& 2 \hat s (1-v_l^2) Re [F H^*] + 2 m_B^2 \hat s (1-r_\phi)
(1-v_l^2) Re[G H^*]
\nn\\
&+&\frac{2}{3}(3-v_l^2)(r_\phi+\hat s-1) Re[B C^*] \biggr]
+\frac{1}{3r_\phi} m_B^2 \biggr[(\lambda +12 r_\phi \hat s)(3-v_l^2) |B|^2 +
\lambda m_B^4\Big[ \lambda (3-v_l^2)\nn\\
&-&3 \hat s (\hat s -2 r_\phi-2)(1-v_l^2)\Big]|G|^2+
\Big(\lambda(3-v_l^2)+24 r_\phi \hat s v_l^2 \Big)|F|^2 \biggr]\;.
\eea
Another observable is the lepton forward backward asymmetry ($A_{FB}$),
which is also a very powerful tool for looking  new physics. The
position of the zero value of $A_{FB}$ is very sensitive to
the presence of new physics.
The normalized forward-backward asymmetry
is defined as
\bea
A_{FB}(s) &= &\frac{\ds{\int_0^1 \frac{d^2 \Gamma}{d \s1 d \cos \theta}
d \cos \theta-\int_{-1}^0 
\frac{d^2 \Gamma}{d \s1 d \cos \theta}d \cos \theta}}
{\ds{\int_0^1 \frac{d^2 \Gamma}{d \s1 d \cos \theta}d
\cos \theta +\int_{-1}^0 
\frac{d^2 \Gamma}{d \s1 d \cos \theta}d \cos \theta}}\nn\\ \nn\\
&=&\frac{G_F^2 \alpha^2}{2^{12} \pi^5 m_B} |V_{tb}V_{ts}^*|^2\
\frac{ 8 m_B^4~ \lambda~ v_l^2~  \hat s~ 
\Big(Re[B E^*]+Re[A F^*] \Big) }{d \Gamma/ds}\;, \label{fb}
\eea
where $\theta $ is the angle between the directions of
$l^+$ and $B_s$ in the rest frame of the lepton pair.

For numerical evaluation we use 
the form factors calculated in the LCSR \cite{ball}
approach, where the $q^2$ dependence of various form factors are given 
by simple fits as
\bea
f(q^2) &=& \frac{r_2}{1-q^2/m_{fit}^2}\;,~~~
~~~~~~~~~~~~~~~~~~~~~~~~~~({\rm for}~~A_1,~T_2)\nn\\
f(q^2) &= & \frac{r_1}{1-q^2/m_{R}^2}+  
\frac{r_2}{1-q^2/m_{fit}^2}\;,
~~~~~~~~~~({\rm for}~~V,~ A_0,~ T_1)\nn\\
f(q^2) &= & \frac{r_1}{1-q^2/m_{fit}^2}+  \frac{r_2}{(1-q^2
/m_{fit}^2)^2}\;,
~~~~~~({\rm for}~~A_2, ~\tilde T_3)\;.
\eea
The values of the parameters $r_1$, $r_2$, $m_R$ and  $m_{fit}$
are taken from \cite{ball}. The uncertainties occur in these fitted 
results are also given in \cite{ball}.
The form 
factors $A_3$ and $T_3$ are given as
\bea
A_3(q^2) &=& \frac{m_B+m_V}{2 m_\phi}A_1(q^2) - 
\frac{m_B-m_\phi}{2 m_\phi} A_2(q^2)\;,\nn\\
T_3(q^2) &=& \frac{m_B^2-m_\phi^2}{q^2}\Big(\tilde{ T}_3(q^2)-T_2(q^2)\Big).
\eea
The particle masses, lifetime of $B_{s}$ meson and the weak mixing angle 
as $\sin^2 \theta_W$ are taken from \cite{pdg}. The quark masses
(in GeV) used are
$m_b$=4.6, $m_c$=1.5  and the CKM matrix elements as
$V_{tb} V_{ts}^*=0.04$ .

In order to see the effect due to the uncertainties in the form factors, 
we plot the differential branching ratio 
(\ref{lp1}) and the forward backward asymmetry (\ref{fb}) for
$B_s \to \phi~ \mu^+ \mu^- $  in Figure-1. It can be seen
from the figure that due to the uncertainties in the form factors
these distributions deviate slightly from their corresponding
central values in the low $s$ region. However, in the large $s$
region these deviations are highly suppressed and the
zero position in the forward backward asymmetry is insensitive 
to these uncertainties. Therefore, we will not consider the 
effect of these uncertainties for the differential decay 
distributions and the forward backward asymmetries but we will 
incorporate their effects in the total decay rates.

\begin{figure}[htb]
   \centerline{\epsfysize 2.0 truein \epsfbox{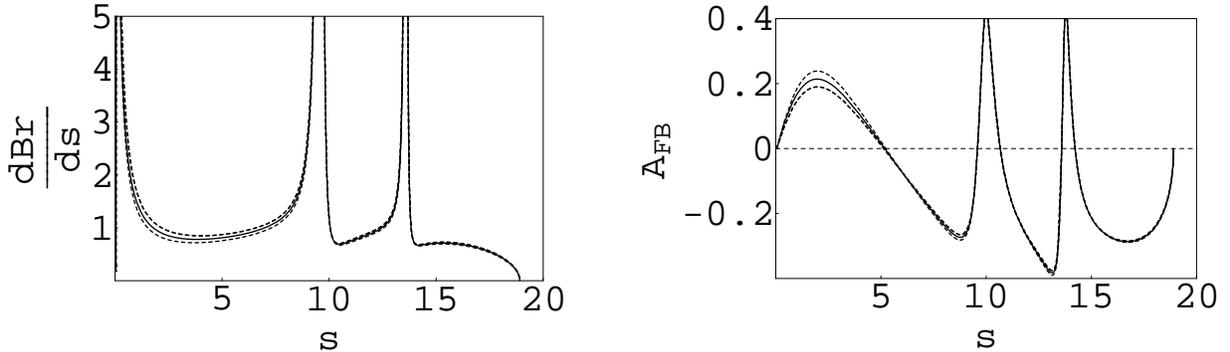}}
 \caption{
 The variation of the differential branching ratio 
$d {\rm Br}/d s$ 
(in units of 
$ 10^{-7})$ and the forward backward asymmetry in the SM 
with $s~({\rm in~GeV^2})$
for the process  $B_s \to  \phi \mu^+ \mu^- $ . The solid line denotes the 
result using the central values of the form factors and
the dashed lines represent the effects due to the uncertainties
in these parameters.}
\end{figure}

Next, to see the effect of the UED, we plot the differential decay 
rates (\ref{lp1}) and the forward backward asymmetries (\ref{fb}) for
$B_s \to \phi~ l^+ l^- $  against $s$, as depicted in Figures 2
and 3. It can be seen from these figures that
there is considerable enhancement in the decay rates due to the
KK contributions for $1/R=200$ GeV. 
The zero position of the forward backward asymmetry
$A_{FB}$ shifts towards the left due to the NP effect. This 
shifting is also more prominent for $1/R=200$ GeV. Therefore,
its experimental determination would constrain the parameter
$1/R$.


\begin{figure}[htb]
   \centerline{\epsfysize 2.0 truein \epsfbox{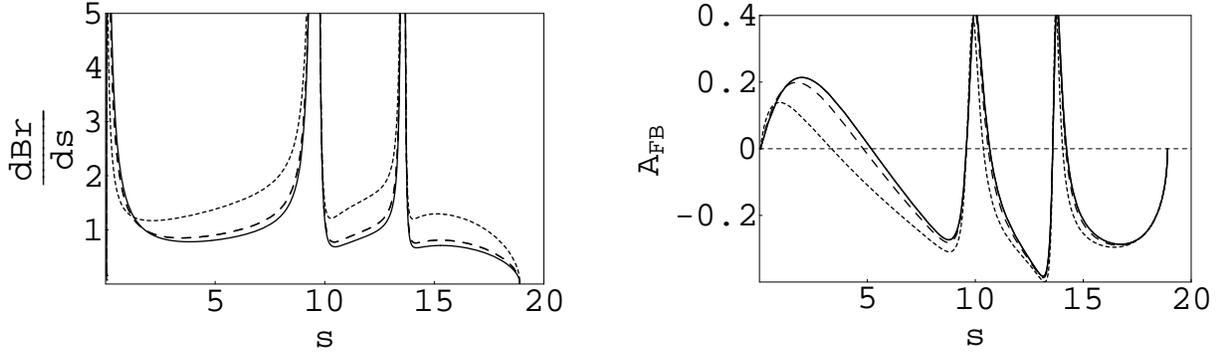}}
 \caption{
 The variation of the differential branching ratio $d {\rm Br}/d s$ 
(in units of 
$ 10^{-7})$ and the forward backward asymmetry with $s~({\rm in~GeV^2})$
for the process  $B_s \to  \phi \mu^+ \mu^- $. The solid line denotes the 
SM result, the dashed line represents the
contribution from UED model with $1/R$ = 200 GeV and the long-dashed line
for $1/R$=500 GeV.}
  \end{figure}

\begin{figure}[htb]
   \centerline{\epsfysize 2.0 truein \epsfbox{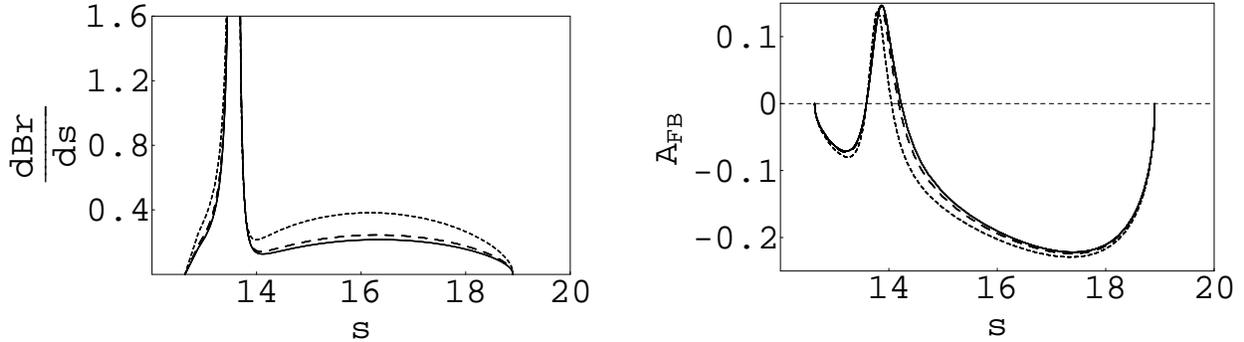}}
 \caption{Same as Figure-2 for the process $B_s \to  \phi \tau^+ \tau^- $ }
  \end{figure}

\begin{figure}[htb]
   \centerline{\epsfysize 3.0 truein \epsfbox{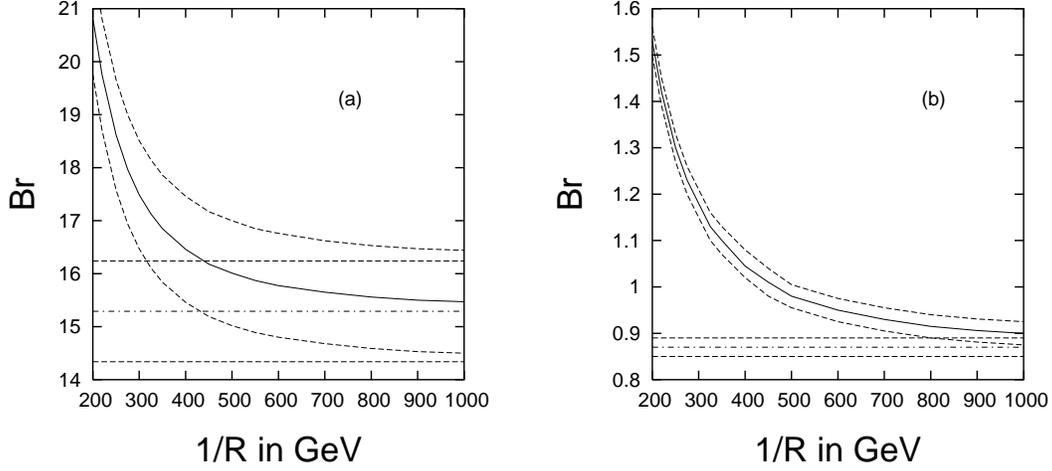}}
 \caption{The variation of the total branching ratio 
in units of $10^{-7}$ with $1/R$.
The solid line represents the central value while the dashed 
lines are due to the uncertainties in the form factors. 
Figure-(a) represents the  $B_s \to  \phi \mu^+ \mu^- $ and
figure (b) $B_s \to  \phi \tau^+ \tau^- $ process. 
The horizontal dot-dashed lines in both the figures represent the 
central values in the SM and the dashed lines are the 
corresponding uncertainties. }
  \end{figure}

\begin{table}
\begin{center}
\caption{The branching ratios 
for various rare semileptonic (in units of $10^{-7}$)
 and radiative leptonic (in units of $10^{-9}$)
$B_s$ decays, where
${\rm {Br}^{SM}}$ represents the SM branching ratio and 
${\rm {Br}|_{1/R=200}}$ is 
the branching ratio with KK contributions for $1/R=200$ GeV
and ${\rm {Br}|_{1/R=500}}$ for $1/R=500$ GeV. The errors are
due to the uncertainties in the form factors. }
\vspace*{0.3 true cm}
\begin{tabular}{|c|cc|c c|cc|cc|}
\hline
\hline
Decay process & & $ {\rm{{Br}^{SM}}} $ && ${\rm {Br}|_{1/R=200}} $
 && ${\rm {Br}|_{1/R=500}}$ && Expt.\cite{ref4}  \\
\hline
$B_s \to \phi \mu^+ \mu^-$ && $15.29 \pm 0.95$  
&& 20.82 $\pm$ 1.06 && 16.01$\pm$ 0.99 && $< 41 $  \\
$B_s \to \phi \tau^+ \tau^-$ && 0.87$\pm$0.02  && 
1.53 $\pm$ 0.03 && 0.98$\pm$0.025 &&  -  \\
$B_s \to \mu^+ \mu^- \gamma$ && 3.05$\pm$0.50  && 
4.45$\pm$ 0.69 && 3.27$\pm$0.53 &&  -  \\
$B_s \to \tau^+ \tau^- \gamma $ && 13.93$\pm$0.09  
&& 24.93$\pm$0.14 && 15.77 $\pm$ 0.10&&  -  \\
\hline
\hline
\end{tabular}
\end{center}
\end{table}

We now proceed to calculate the  total decay rate for $B_s \to
\phi~ l^+ l^-$. It should be noted that the long distance 
contributions arise from the real $\bar c c$ resonances with the
dominant contributions coming from the low
lying resonances $J/\psi$ and $\psi'$. In order to
minimize the hadronic uncertainties it is necessary to eliminate 
the backgrounds coming from the resonance regions. 
This can be done by using the
following  cuts for
$B_s \to  \phi \mu^+ \mu^- $ process as \cite{d0}
\begin{eqnarray*}
2.9~{\rm GeV} <
m_{\mu^+ \mu^-}<3.3~{\rm GeV}\;,~~{\rm and}~~
 3.6~{\rm GeV} <
m_{\mu^+ \mu^-}<3.8~{\rm GeV}\;,
\end{eqnarray*}
and for $B_s \to \phi \tau^+ \tau^-$ process $
 m_{\psi^\prime}-0.08<m_{\tau^+ \tau^-}<m_{\psi^\prime}+0.08\;.$

Using these veto windows we obtain the branching ratios for
the semileptonic rare $B_s$ decays which are given in Table-1.
It is seen from the table that the branching ratios obtained
in the ACD model  are enhanced from the
corresponding SM values for lower $1/R$ value. The variation of the 
total decay rates with $1/R$ are depicted in Figure-4. Thus,
observation of these decay modes can be used to
constrain the compactification radius $R$. As seen from the figure
for large $1/R$ the standard model results are recovered.   

\subsection{ $B_s^0 \to l^+ l^- \gamma$ process}

Now let us consider the radiative dileptonic decay modes
$B_s \to l^+ l^- \gamma $, which are also very sensitive to the
existence of new physics beyond the SM. Due to the presence
of the photon in the final state, these decay modes are free from
helicity suppression, but they are further suppressed by a
factor of $\alpha $. However, in spite of this
$\alpha $ suppression,  the radiative leptonic decays $B_s \to 
l^+ l^- \gamma $, $l=(\mu, \tau)$  have comparable decay rates to that of
purely leptonic ones. 

The matrix element for the decay $B_s \to l^+ l^- \gamma$ can
be obtained from that of the  $B_s \to l^+ l^-$ one  by attaching the photon
line to any of the charged external fermion lines. In order to
calculate the amplitude, when the photon is
radiated from the initial fermions (structure dependent (SD) part), we
need to evaluate the  matrix elements of the quark currents present
in (\ref{ham}) between the emitted  photon and the initial $B_s$
meson. These matrix elements can be obtained by considering the
transition of a $B_s$ meson to a virtual photon with momentum $k$.
In this case the form factors depend on two variables, i.e., $k^2$ (the
photon virtuality) and the square of momentum transfer $q^2=(p_B-k)^2$.
By imposing gauge invariance, one can obtain several relations
among the form factors at $k^2=0$. These relations can be used 
to reduce the number of independent form factors for the transition of 
the $B_s$ meson to a real photon. Thus, the matrix elements for $B_s
\to \gamma$ transition, induced by vector, axial-vector, tensor and 
pseudo-tensor currents can be parametrized as \cite{kruger}
\bea
\langle \gamma(k, \ve)|\bar s \gamma_\mu \gamma_5 b|B_s(p_B) \rangle
&=& ie \left [ \ve_\mu^* (p_B\cdot k) -(\ve^* \cdot p_B) k_\mu \right ]
\frac{F_A}{m_{B}}\;,\nn\\
\langle \gamma(k, \ve)|\bar s \gamma_\mu  b|B_s(p_B) \rangle
&=& e\epsilon_{\mu \nu \alpha \beta} \ve^{*\nu} p_B^\alpha~ k^\beta
\frac{F_V}{m_{B}}\;,\nn\\
\langle \gamma(k, \ve)|\bar s \sigma_{\mu \nu} q^\nu 
\gamma_5b|B_s(p_B) \rangle
&=& e \left [ \ve_\mu^* (p_B\cdot k) -(\ve^* \cdot p_B) k_\mu \right ]
F_{TA}\;,\nn\\
\langle \gamma(k, \ve)|\bar s \sigma_{\mu \nu} q^\nu b|B_s(p_B) \rangle
&=& e\epsilon_{\mu \nu \alpha \beta} \ve^{*\nu} p_B^\alpha~ k^\beta
F_{TV}\;,
\eea
where $\varepsilon$ and $k$ are the polarization vector and the
four-momentum of photon, $p_B$ is the momentum of initial $B_s$
meson and $F_i$'s are the various form factors.  

Thus, the matrix element describing the SD part takes the form
\bea
{\cal M}_{SD} &=& \frac{\alpha^{3/2}G_F}{\sqrt{2 \pi}}~ 
V_{tb}V_{ts}^*
 \biggr\{ \epsilon_{\mu \nu \alpha \beta}
\varepsilon^{* \nu} p_B^\alpha~ k^\beta\Big(A_1~ \bar l \gamma^\mu l
+A_2~ \bar l \gamma^\mu \gamma_5 l \Big)\nn\\
&+&
i\Big( \varepsilon_\mu^*(k \cdot p_B)-(\varepsilon^* \cdot p_B) k_\mu
\Big)\Big(B_1~ \bar l \gamma^\mu l
+B_2~ \bar l \gamma^\mu \gamma_5 l \Big)\biggr\}\;,
\label{sd}
\eea
where
\bea
A_1&=& 2 C_7 \frac{m_b}{q^2}F_{TV}+C_9 \frac{F_V}{m_{B}}\;,~~~
~~~~~~~~~~A_2=C_{10}\frac{F_V}
{m_{B}}\;,\nn\\
B_1&=& -2C_7\frac{m_b}{q^2} F_{TA}-C_9 \frac{F_A}{m_{B}}\;,~~~
~~~~~~~~~B_2=-C_{10} \frac{F_A}{m_B}\;.\label{ff}
\eea
The form factors $F_V$ and $F_A$ have been calculated
within the dispersion approach \cite{new1}.
The  $q^2$ dependence of the
form factors are given as \cite{kruger}
\be
F(E_\gamma)= \beta \frac{f_{B_s} m_{B}}{\Delta+ E_\gamma}\;,
\label{ib}
\ee
where $E_\gamma$ is the photon energy, which is related to the 
momentum transfer $q^2$ as
\be
E_\gamma= \frac{m_{B}}{2}\left (1- \frac{q^2}{m_{B}^2} \right )\;.
\ee
The values of the parameters $\beta $ and $\Delta$
 are given in Table-2. The same ansatz 
(\ref{ib}) has also been assumed for the form factors 
$F_{TA}$ and $F_{TV}$. The decay constant of the $B_s$
meson is not yet well known because the pure leptonic
decays of $B_s$ meson (i.e., $B_s \to l^+ l^-$) from which
it could be extracted are highly suppressed in the SM as they
occur at one one-loop level. Using QCD Sum rule approach,
its value is found to be $f_{B_s}=236 \pm 30$ MeV \cite{new2}
and $f_{B_s}=244 \pm 21$ MeV \cite{new3}. The Lattice QCD result
gives $f_{B_s}=242 \pm 9 \pm 34$ MeV \cite{new4}. Therefore, in this analysis
we use the value $f_{B_s}=240$ MeV.
\begin{table}
\begin{center}
\caption{The parameters for $B \to \gamma$ form factors.}
\vspace*{0.3 true cm}
\begin{tabular}{|c|cc|c c|cc|cc|}
\hline
\hline
Parameter & & $ F_V $ && $ F_{TV}$ && $F_A$ && $F_{TA}$ \\
\hline
$\beta ({\rm GeV}^{-1})$ && 0.28  && 0.30 && 0.26 && 0.33 \\
$\Delta ({\rm GeV})$ && 0.04  && 0.04 && 0.30 && 0.30 \\
\hline
\hline
\end{tabular}
\end{center}
\end{table}

When the photon is radiated from the outgoing lepton pairs, the
internal bremsstrahlung  (IB) part, the matrix
element is given as 
\be
{\cal M}_{IB} = \frac{\alpha^{3/2}G_F}{\sqrt{ 2 \pi}}~
V_{tb}V_{ts}^*~  f_{B_s}~ m_l~ C_{10}
\biggr[ \bar l \left ( \frac{\not\!{\varepsilon}^* {\not\!{p}}_{B}}{
p_+ \cdot k}-\frac{{\not\!{p}}_{B}\not\!{\varepsilon}^* }{
p_- \cdot k} \right )\gamma_5 ~l \biggr]\;.
\ee
Thus, the total matrix element for the $B_s \to l^+ l^- \gamma $ process
is given as
\be
{\cal M}={\cal M}_{SD}+{\cal M}_{IB}\;.
\ee
The differential decay width of the $B \to l^+
l^- \gamma $ process, in the rest frame of $B_s$ meson is given as 
\be
\frac{d \Gamma}{d s}= \frac{G_F^2 \alpha^3}{2^{10} \pi^4}~ |V_{tb} 
V_{ts}^*|^2~
m_{B}^3~ \Delta_1\;,\label{lp}
\ee
where
\bea
\Delta_1 &=& \frac{4}{3} m_{B}^2 (1- \hat s)^2 v_l \Big((\hat s+2r_l)(|A_1|^2
+|B_1|^2)+(\hat s-4 r_l)(|A_2|^2+|B_2|^2 \Big )\nn\\
&-& 64~ \frac{f_{B_s}^2}{m_{B_s}^2} \frac{r_l}{1- \hat s}~ C_{10}^2~\Big(
(4r_l-\hat s^2 -1) \ln
\frac{1+v_l}{1-v_l}+2 \hat s~ v_l\Big)\nn\\
&-& 32~r_l(1-\hat s)^2~ f_{B_s} {\rm Re}\Big( C_{10} A_1^* \Big),
\eea
with  $s=q^2$, $\hat s= s/m_{B}^2$, $r_l=m_l^2/m_{B}^2$,
$v_l=\sqrt{1- 4 m_l^2/q^2}$. The physical region of $s$ is 
$4 m_l^2 \leq s \leq m_{B}^2 $.

The forward backward asymmetry is given as
\bea
A_{FB} &=& \frac{1}{\Delta_1} \biggr[2 m_{B}^2 \hat s(1-\hat s)^3 v_l^2~
{\rm Re}\Big(
A_1^* B_2+B_1^* A_2\Big)\nn\\
&+&32~ f_{B_s}~r_l (1-\hat s)^2 \ln\left (\frac{4r_l}{\hat s} \right )
{\rm Re}\Big(C_{10}B_2^*\Big)\biggr]\;.\label{fb1}
\eea 
Again to visualize the effects due to the uncertainties in 
the form factors (assuming it to at the level of $10\%$) 
we first plot the differential  decay
distribution (\ref{lp}),
and the forward backward asymmetry (\ref{fb1}) for $B_s \to 
\mu^+ \mu^- \gamma$ in Figure-5. From the figure it can be seen that
these uncertainties can affect only the differential decay rate but not
the forward backward asymmetry. Furthermore, it is found that these
effects are significant only for $B_s \to \mu^+ \mu^- \gamma$ and
not for $B_s \to \tau^+ \tau^- \gamma$. Next, we would like to
see the effect of UED in these differential distributions. For 
this purpose, we consider only the central values of the form factors,
$f_{B_s}$=0.24 GeV, $\alpha$=1/128 and plot
 plot the dilepton mass spectrum (\ref{lp}),
and the forward backward asymmetries (\ref{fb1}) for $B_s \to l^+ l^- \gamma$ 
decays which are shown in Figures-6 and 7.  From these figures,
we see that the branching ratio for $B_s \to l^+ l^- \gamma $ enhanced
significantly from their corresponding SM values. However, the forward 
backward asymmetry is reduced slightly from the corresponding SM value
for the $B_s \to \mu^+ \mu^- \gamma$ process and there is a backward 
shifting of the zero position. For  $B_s \to \tau^+ \tau^- \gamma$ process
there is no significant change in the asymmetry distribution.
\begin{figure}[htb]
   \centerline{\epsfysize 2.0 truein \epsfbox{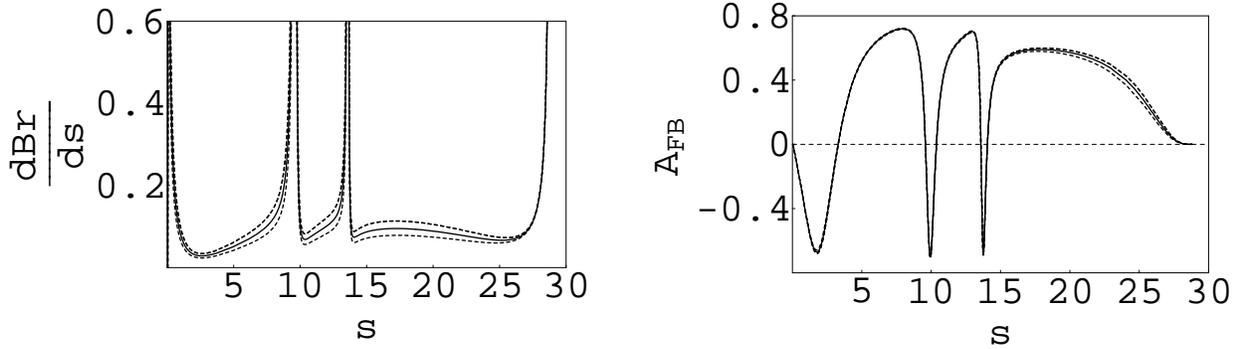}}
 \caption{
 The differential branching ratio (\ref{lp}) (in units of $10^{-9}$) and 
the forward backward 
asymmetry ($A_{FB}$) for the process  
$B_s \to   \mu^+ \mu^- \gamma $ in the standard model where the
solid line represents the central value and the dashed lines 
represent the uncertainties due to the form factors.
 }
  \end{figure}
\begin{figure}[htb]
   \centerline{\epsfysize 2.0 truein \epsfbox{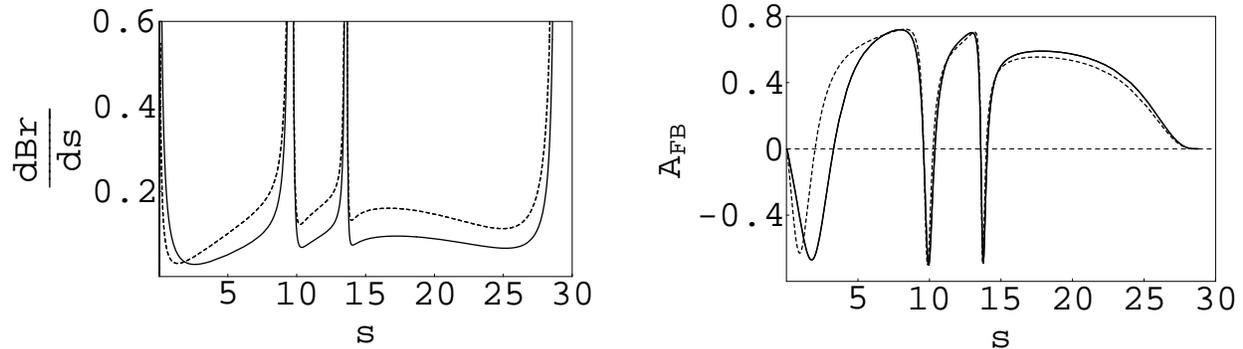}}
 \caption{
 The differential branching ratio (\ref{lp}) (in units of $10^{-9}$) and 
the forward backward 
asymmetry ($A_{FB}$) for the process  
$B_s \to   \mu^+ \mu^- \gamma $ in the standard model and in
the  ACD model. The solid line denotes the 
SM result, the dashed line represents the
contribution from UED model with $1/R$ = 200 GeV. }
  \end{figure}
\begin{figure}[htb]
   \centerline{\epsfysize 2.0 truein \epsfbox{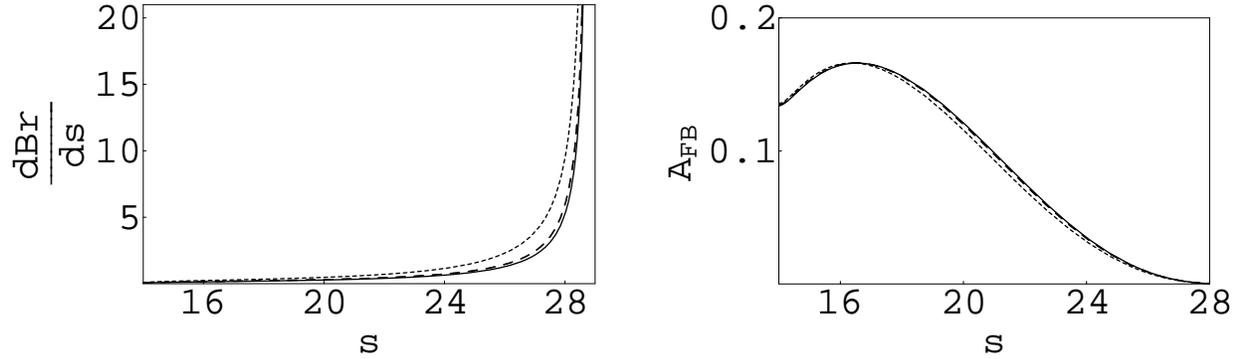}}
 \caption{Same as Figure-7 for the process  
$B_s \to   \tau^+ \tau^- \gamma $. The additional long-dashed line
is for $1/R$=500 GeV.}
  \end{figure}

To obtain the branching ratios it is necessary
to eliminate the background due to the resonances  $J/\psi 
(\psi^\prime)$
with $J/\psi(\psi^\prime) \to l^+ l^- $. We use the
following  veto
windows to eliminate these backgrounds
\begin{eqnarray*}
B_s \to \mu^+ \mu^- \gamma :&&m_{J/\psi}-0.02<
m_{\mu^+ \mu^-}<m_{J/\psi}+0.02;\nn\\
:&&
 m_{\psi^\prime}-0.02<m_{\mu^+ \mu^-}<m_{\psi^\prime}+0.02 \nn\\
B_{s} \to  \tau^+ \tau^- \gamma:&&
m_{\psi^\prime}-0.02<m_{\tau^+ \tau^-}<m_{\psi^\prime}+0.02 \;.
\end{eqnarray*}
Furthermore, it should be noted that the 
$|{\cal M}_{IB}|^2$ has infrared singularity
due to the emission of soft photon. Therefore, to obtain the branching ratio,
we impose a cut on the photon energy, which will correspond to the
experimental cut imposed on the minimum energy for the detectable photon.
Requiring the photon energy to be larger than 25 MeV, i.e.,
$E_\gamma \geq \delta~ m_{B_s}/2$, which corresponds to
$s \leq m_{B_s}^2(1- \delta)$, and therefore, we set 
the cut $\delta \geq 0.01 $.

Thus, with the above defined veto windows and the infrared cutoff parameter,
 we obtain the branching ratios as shown in Table-1
which are enhanced from their SM values.
It should be mentioned that the $B_s^0 \to \tau^+ \tau^- \gamma$
could be observable in the Run II of Tevatron.

\section{Conclusion}

It is now widely believed that the scenario with extra
dimensions is a strong contender to reveal physics beyond the SM 
and has received considerable attention in the literature. In view of this
anticipation it is also worthwhile to study its implications in the b-sector.
In this paper, therefore,  we have studied the rare semileptonic decay mode 
 $B_s \to \phi~ l^+ l^-$ and the radiative leptonic
 $B_s \to l^+ l^- \gamma $ in the model with a single extra
dimension. The branching ratios for various decay modes are
 found to be
larger than their corresponding SM values but they lie
within the present experimental upper limit.
 Furthermore, the zero point of the forward backward 
asymmetries for the decays under consideration are shifted to the left
and the change is found to be sensitive to the inverse compactification
radius. In future, with more intensive data, the UED scenario will
be subjected to more stringent tests and in turn will enrich us with a
better understanding of the flavor sector.

{\bf Acknowledgments}
The work of RM was
partly supported by the Department of Science and Technology,
Government of India, through Grant No. SR/S2/HEP-04/2005.

\end{document}